\documentclass[final,5p,times,twocolumn]{elsarticle}

\usepackage{amssymb}

\makeatletter
\newcommand{\verbatimfont}[1]{\def\verbatim@font{#1}}%
\newcommand{\footurl}[1]{\footnote{\texttt{\url{#1}}}}

\newcommand{\ray}{\texttt{[Ray~Casting: Compositing]}} 
\newcommand{\kb}{\texttt{[Kernel-based]}}              

\newcommand{\Xeon}{Xeon} 
\newcommand{\intel}{Intel}

\newcommand{\sng}{SuperMUC-NG} 

\newcommand{\py}{{Python}}   
\newcommand{\mesa}{{Mesa}}   
\newcommand{\visit}{VisIt}      
\newcommand{\ospray}{OSPRay} 
\makeatother
\usepackage{url}

\usepackage{xcolor}
\usepackage[normalem]{ulem}

\journal{Parallel Computing}
\begin{document}
\begin{frontmatter}

\title{Visualizing the world's largest turbulence simulation}

\author[lrz]{Salvatore Cielo}
\author[lrz]{Luigi Iapichino}
\author[intel]{Johannes G{\"u}nther}
\author[uca]{Christoph Federrath}
\author[lrz]{Elisabeth Mayer}
\author[lrz]{Markus Wiedemann}

\address[lrz]{Leibniz Supercomputing Centre of the Bavarian Academy of Sciences and Humanities (LRZ), Garching b. M{\"u}nchen, Germany}
\address[intel]{Intel Corporation, Feldkirchen, Germany}
\address[uca]{Research School of Astronomy and Astrophysics, Australian National University, Canberra, ACT~2611, Australia}

\begin{abstract}
   In this exploratory submission we present the visualization of the 
   largest interstellar turbulence simulations ever performed, unravelling key astrophysical processes concerning the formation of stars and the relative role of magnetic fields. The simulations, including pure hydrodynamical (HD) and magneto-hydrodynamical (MHD) 
   runs, up to a size of $10048^3$ grid elements, were produced on the supercomputers of the Leibniz Supercomputing Centre and visualized using the hybrid parallel (MPI + TBB) ray-tracing engine \ospray\ associated with \visit. Besides revealing features of turbulence with an unprecedented resolution, the visualizations brilliantly showcase the stretching-and-folding mechanisms through which astrophysical processes such as supernova explosions drive turbulence and amplify the magnetic field in the interstellar gas, and how the first structures, the \emph{seeds} of newborn stars are shaped by this process.
   
\end{abstract}


\end{frontmatter}


\section{The link between turbulence and star formation in the Universe}

\begin{figure*}[h]
      \centering
      \includegraphics[width=\textwidth]{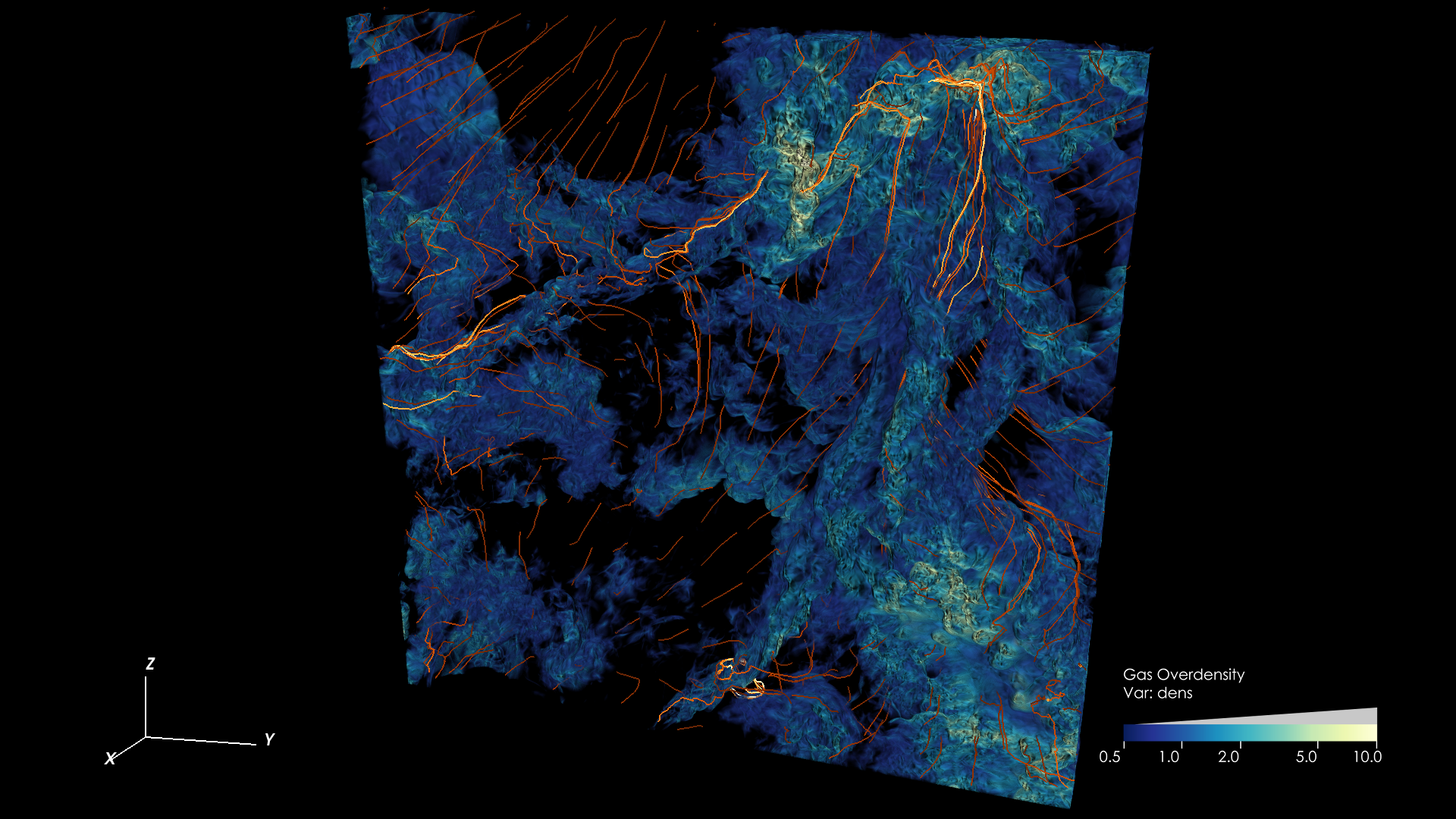}
       \caption{Sample ray-tracing plus velocity streamlines rendering of the $1152^3$ HD simulation, using \visit{} and \ospray{} on SuperMUC-NG. The displayed data correspond to a slab with a volume of $10\%$ of the full data cube.}
     \label{f-front}
 \end{figure*}
  
Turbulent flows are a key feature for characterizing the dynamics of physical systems 
in several domains, including weather modelling, industry and astrophysics. While turbulence gives rise to long-term unpredictable systems, 
its effects have been characterized in great detail. However, for quantitative predictions, statistical and numerical methods are essential tools.

Turbulence is ubiquitous in astrophysical fluids; just to name a few examples, it plays an important role in accelerating cosmic ray particles and amplifying magnetic fields in clusters of galaxies, and in the flame propagation mechanism of thermonuclear supernovae. Coming to the main topic of this work, turbulence is an essential ingredient for modeling the physics of the interstellar medium (ISM), the astrophysical site where stars are formed.


Numerical simulations of turbulent fluids in the ISM provide insight into the statistical properties of the flow and are necessary for a quantitative characterization of star formation. While in principle these simulations contain nothing but fundamental physics and hydrodynamics (HD), the multi-scale distribution of energy, the diffuse high-velocity regions and the chaotic flows of these computations make domain-optimization techniques (such as the use of adaptive meshes, or grids different than Cartesian) rather ineffective. Thus, such simulations have to rely purely on numerical and solver optimizations, and on an efficient parallelization scheme for running on large, modern supercomputing architectures.

Turbulent models of the ISM also show a rich pool of emergent physical effects \cite{fed18}.
The qualitative picture presented by observations with modern telescopes such as the Atacama Large Millimetre/Sub-millimetre Array (ALMA) in the Chilean desert is that turbulence is driven on large scales, e.g.~by shocks and feedback from nearby or newly born stars. These stirring mechanisms have an episodic character, yet the resulting supersonic turbulent motions are actively sustained for time scales longer than the typical interval between driving events.
The kinetic energy of the turbulent motions is then transferred to smaller length scales in the cascade process, until turbulence is dissipated by viscous effects on small scales. In this process, at a typical length called \emph{sonic scale} a transition from supersonic to subsonic turbulence occurs, marking a corresponding change of the statistical properties of the turbulent energy spectrum.

Thus, a simulation capable of resolving this all-important \emph{sonic scale} 
is a necessary first step in understanding the structure of star-forming regions. Studying this problem is computationally very challenging because it requires enormous spatial resolution, resulting in a sufficiently large dynamical range.
\begin{figure}[h]
      \centering
      \includegraphics[width=0.495\textwidth]{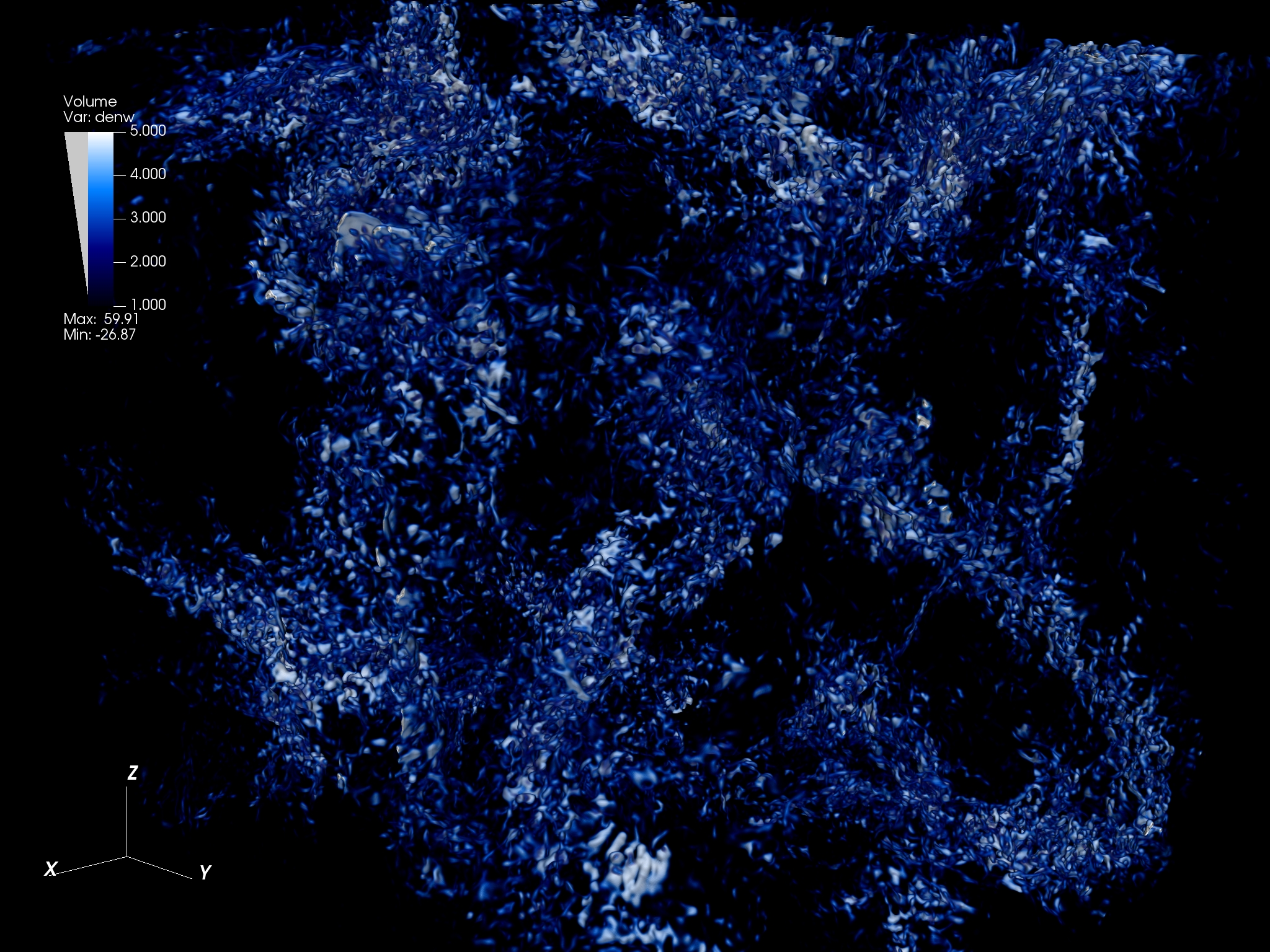}
       \caption{Volume rendering of the gas density structures associated with the sonic scale (i.e.~regions with the gas velocity close to the sound speed) in the HD simulation with grid resolution of $1152^3$.}
     \label{f-sonic}
 \end{figure}
  
Finally, most observed star-forming regions have clear indications of the presence of magnetic fields. This additional physical ingredient changes the spectrum of turbulence both quantitatively (shifting the energy distribution) and qualitatively (giving rise to possible self-amplifying dynamo effects). Modern codes can deal effectively with magnetic fields in the magneto-hydrodynamics (MHD) framework, but this roughly doubles the computational requirements of already highly demanding simulations.

\subsection{The largest turbulence simulations to date}

We present a visualization of the largest sets of HD and MHD turbulence simulations ever realized, with up to $10048^3$ resolution elements in a static grid (see Figure \ref{f-front}, based on a smaller test run). This unprecedented resolution allows for a dynamic range of four orders of magnitude in length scale.
The simulations were produced with the FLASH code \cite{fryxell2000FLASH}, a public grid-based MHD code. 
In order to model the injection of supersonic turbulence,
FLASH uses a stochastic, Fourier-based driving method (see e.g.~\cite{frk10}).

The detailed results of the HD simulation were presented in \cite{federrath2019sonicscale}; see also \cite{fki16} for a technical overview. A quantitative characterization of the gas Mach number as a function of spatial scale (the so-called \emph{structure function}) has revealed excellent agreement with theoretical models both in the supersonic and subsonic regime.
For the first time, the sonic scale is resolved as the transition between them (Figure \ref{f-sonic}). This in turns allows us
to infer the width distribution of filamentary structures in star-forming regions and ultimately the critical density for the formation of stars.

As for the MHD simulation, besides technical works describing the code performance \citep{cielo2019modernization}, the present submission displays its outputs for the first time. Its scientific results will be presented in forthcoming publications.  

\section{Computational details}

\subsection{Code optimization and performance}
\begin{figure}[t]
    \centering
    \includegraphics[width=\columnwidth]{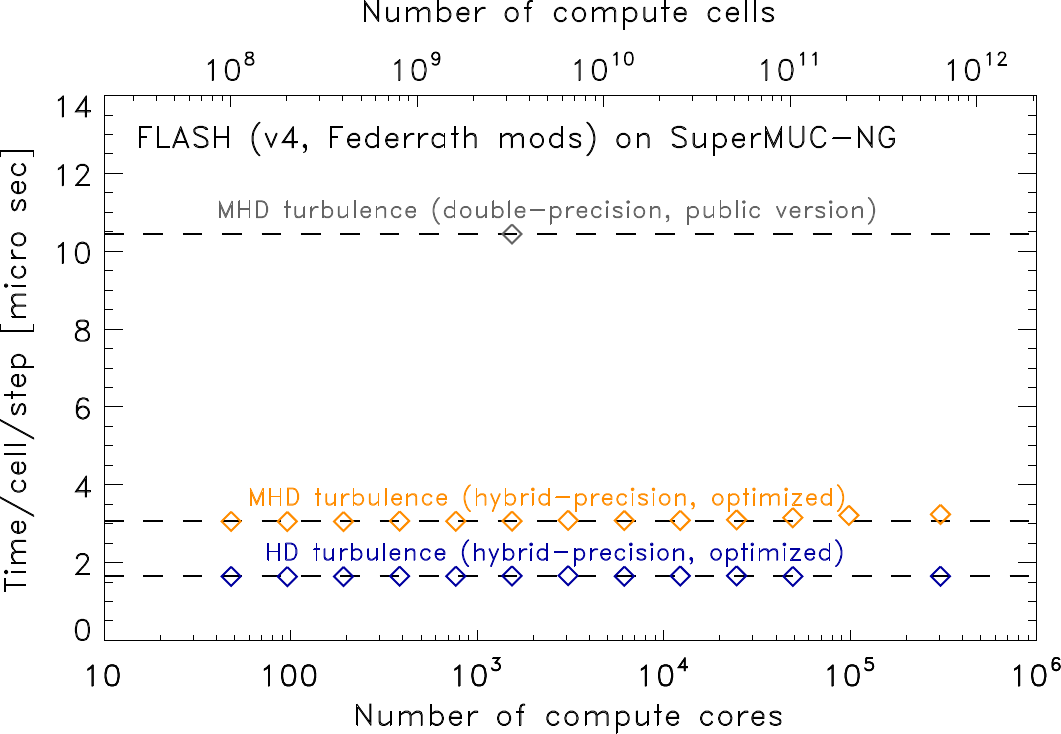}
    \caption{Weak scaling test of FLASH on SuperMUC-NG. Gray, orange and blue symbols correspond to MHD simulations run with the public code version, and MHD and HD simulations run with the hybrid-precision code version, respectively. The horizontal dashed lines mark the ideal scaling in all three cases. The scaling remains optimal up to full machine size for the runs with the optimized code. MHD introduces a performance overhead with respect to HD, but the code largely benefits from the hybrid-precision scheme.}
    \label{f-flash}
\end{figure}

 The version of FLASH used in this work, based on the public release v.~4, has been optimized to introduce a hybrid single/double precision scheme (\cite{fki16}) that, while preserving the same accuracy of the double-precision case, allows for a $3.6\times$ speedup and a factor $4.1$ lower memory consumption. These numbers refer to the MHD simulations, run on the \emph{SuperMUC-NG} supercomputer at LRZ\footurl{https://doku.lrz.de/display/PUBLIC/SuperMUC-NG}, whose architecture is based on \intel{}\textsuperscript{\textregistered} \Xeon{}\textsuperscript{\textregistered} Platinum 8174 Processors (code-named Skylake). FLASH is capable of scaling up to full machine size (6336 nodes, corresponding to 304,128 compute cores; see Figure \ref{f-flash}). 
 The realization of such large simulations presents severe technical issues, both on the computation and data storage viewpoint: the $10048^3$ HD simulations alone required about $131~$TB memory, over $23$~TB of disk space per snapshot (with more than 100 snapshots produced during the simulation campaign), and $45$ millions CPU-h of computing time.

\subsection{Visualization: \visit{} and \ospray{}}
Visualizing such large data sets comes with enormous challenges in their computation.
Volume renderings of scientific data is traditionally based on the use of hardware accelerators. On the other hand, the recent generations of multi-core processors, such as the Skylake nodes which \sng{} is based on, expose a high degree of parallelism (number of cores per node and large vector registers) which would be beneficial to exploit for visualization purposes. 
Ray-tracing methods are particularly well-suited to run on such architectures, as they present excellent parallel scalability \cite{childs2012visit, wald2017ospray}, and easy access to high-quality illumination aftereffects (e.g. shading and ambient occlusion). In addition, being able to perform visualizations (as well as post-processing) on the same machine where the simulations were run presents obvious advantages, in terms of workflow, data management and hardware usage.

With these considerations in mind, we have built and deployed a custom version of the \visit{} analysis and visualization software for \sng{}.
This version integrates the \intel{} \ospray{} rendering engine\cite{wu2018visitospray}\footurl{https://www.ospray.org}, embodying the \emph{software-defined visualization} concept, optimized for CPU usage without the need of accelerators. \ospray{} uses Threading Building Blocks (TBB) for parallel work sharing, and integrates additional features (e.g. the aforementioned shading and ambient occlusion, or transfer functions pre-integration) which are absent in the standard version of \visit{} \cite{childs2012visit}.

Our customized version including \ospray{} is based on the recent public \visit{} v3.0.0. Considerable work on the source code, not included in the public release, was however necessary for granting \ospray{} full communication across the compute nodes via MPI. This build shows better performance than the standard public version of \visit{} used for reference (v.~2.13.2), based on the \mesa{}  GL and parallelized with a pure MPI scheme. 

\begin{figure}[h!]
    \centering
    \includegraphics[width=\columnwidth]{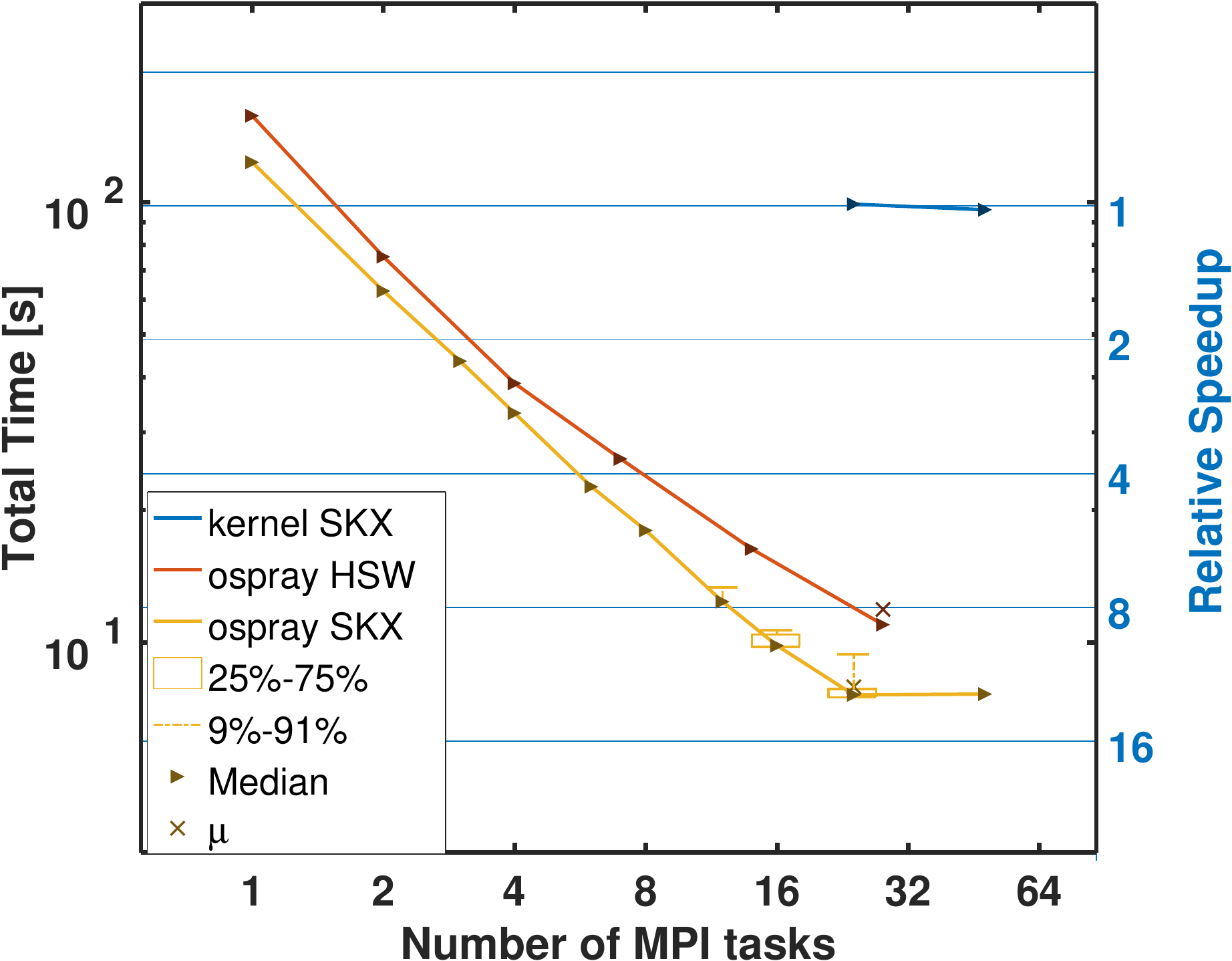}  ~ \\ ~
    \includegraphics[width=\columnwidth]{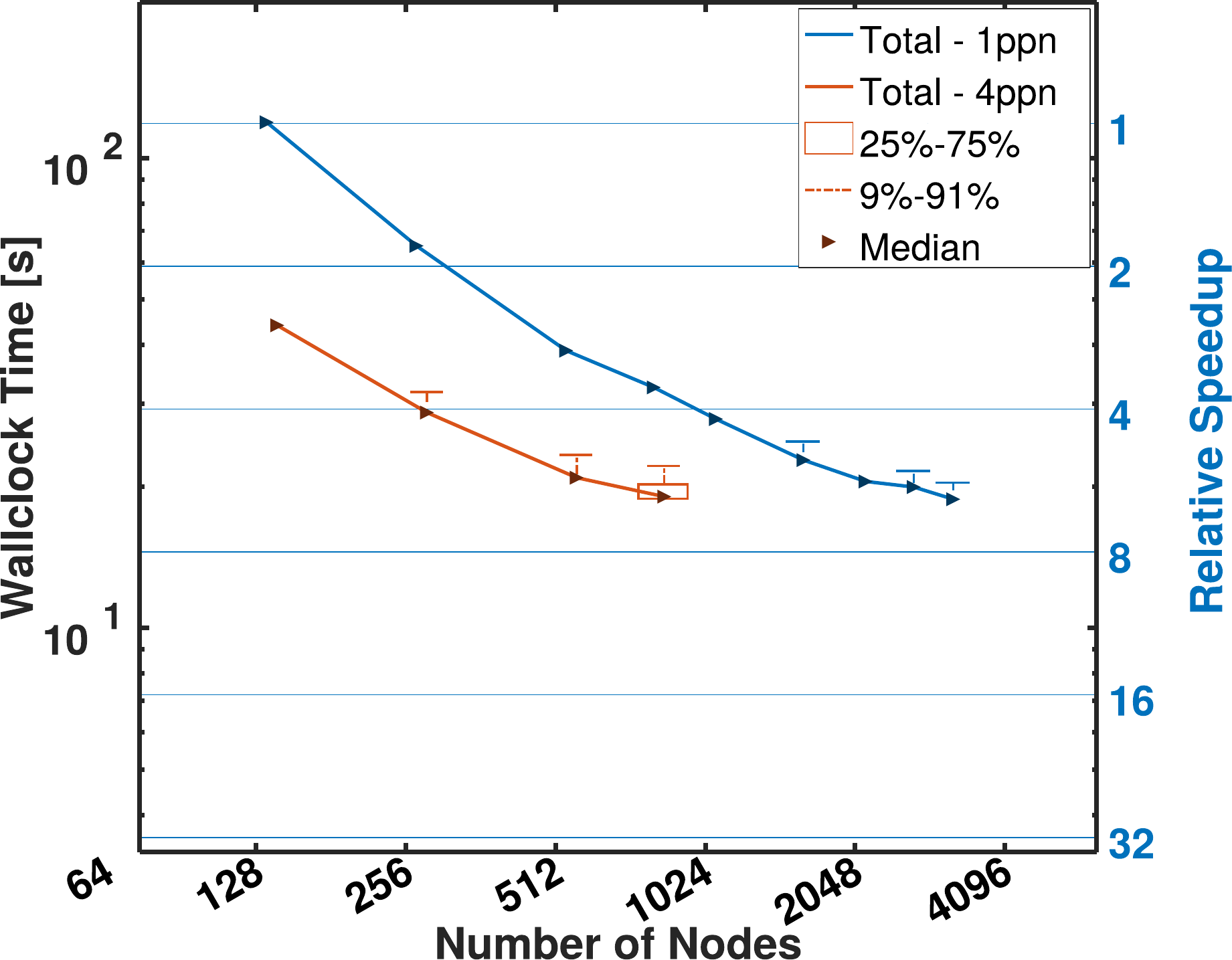}
    \caption{{\sc Upper panel}: \visit{}-\ospray{} node-level scaling behavior over MPI tasks (TBB not shown, but always at work). The yellow and red lines show the scaling for a SuperMUC-NG node (Skylake node) and, for comparison, for a Haswell node (SuperMUC Phase 2). Moreover,
    for the best-scaling algorithm of the standard \visit{} version (kernel-based ray-casting, blue line), two scaling points are shown.  {\sc Lower pane}: \visit{}-\ospray{} strong scaling for tasks of the presented visualization, using one (blue line) or four (red line) MPI tasks per node. The reported time to solution refers to a single snapshot of the \emph{tomography} of the $10048^3$ HD simulation, from $128$ to $3076$ nodes (i.e. one half of the whole \sng{}). Percentile confidence intervals after 20 measures are shown when significant.}
    \label{f-visit_scaling}
\end{figure}

In Figure \ref{f-visit_scaling} we show the scaling behaviour of our visualization strategy. In the upper panel, the node-level scaling is presented, including two reference points from the most scalable and comparable algorithm (according to our previous tests) from the reference \visit{} version, \ray:\kb \cite{childs2012visit}. \ospray{} (yellow line) reaches a speedup of almost $16\times$ on the node. The same algorithm on the \intel{} \Xeon{} E5-2697 v3 architecture (Haswell node from SuperMUC Phase 2, with 28 cores, red line) achieves a nearly comparable speed-up, though with lower absolute performance. The use of \ospray{} improves the time to solution of the rendering on a node (comparison between the blue and yellow line) by $12\times$.

The bottom diagram shows the scaling for a typical task of our submission, the rendering of a single slab of what we call a \emph{tomography} of the gas density, performed on the largest simulation snapshot, the HD with the resolution of $10048^3$ grid cells. This task includes both the creation of a 3D scene and the proper scalable rendering phase\footnote{i.e.~the only task using TBB parallelization.}, but leaving out I/O timings. Satisfactory scaling is achieved up to a considerable fraction of the machine, flattening only when approaching a few thousand compute nodes. 
This feature of the scaling holds when using both one or four MPI tasks per node (blue and red line, respectively).

Given the large size of the visualized data, it is probably worth considering the hypothesis of rendering the volumetric data directly from memory as they are produced by FLASH, a technique commonly know as \emph{on-the-fly} or \emph{in-situ} visualization. The pros of this approach include a time resolution as high as the time stepping of the simulation itself, and the elimination of I/O times. The main cons are instead the non-interactive nature of the task, an overhead that limits the resources available for the simulation, and the necessity to interface simulation and visualization code. Although \visit{} presents this interface capability, this was not a priority for our project. On one hand, the custom version of FLASH we used includes on-the-fly visualization routines, though currently limited to slices and projections; on the other, our main aim is to showcase the efficient visualisation of large amounts of simulation data from storage. Of course \emph{on-the-fly} visualization presents interesting implications, and we may consider it for future projects.

As a final remark we point out a few additional and handy synergies of the software-defined visualization workflow. Firstly, it is possible to access \visit{}-\ospray{} either in interactive GUI-aided sessions, or in batch mode via the SLURM queuing system and \visit{}'s \py{}  scripting, as it was done for the present case. The two methods can at will be conjugated, since the multi-purpose GUI allows us to configure hosts for remote usage, and even to submit batch jobs, a very handy feature on HPC systems. 
Volume plots produced in this way are also easily suited for Virtual Reality (VR) applications, since stereographic rendering can be easily configured. A VR version of our submission is in progress and will be showcased in the LRZ visualization facilities\footurl{https://www.lrz.de/services/v2c_en/}. 
Finally, \visit{} can also be used for data post-processing other than visualization, through queries, histogram plots and other features that also benefit from MPI parallelization.

\section{Movie description and scientific considerations}

The whole visualization consists of raytracing, volumetric plots rendered with \visit{}, using the \ospray{} rendering engine. The only exception is the false-color velocity streamline plot (see below), because this specific kind of visualization is not yet supported by \visit{}-\ospray{}.

The visualization allows the viewer to draw some immediate qualitative conclusions.
The main findings are summarized below.

\subsection{Gas density of the $10048^3$ HD run}

After explaining how the sonic scale is interpreted, we begin the visualization with a \emph{tomography} of the gas overdensity (i.e. the density fluctuations with respect to the mean density).
This tomography consist in showing only the gas contained in a \emph{slab}, whose thickness is $10\%$ of the total box. The slab moves across half of the simulation box, at a fixed final time snapshot of the simulation. With this kind of visualization one can show the minute details of the extreme-resolution data set.
This section of the movie has been rendered with 512 nodes on \sng{}.

The following addition of velocity streamlines (commented in the visualization, and colored according to their path length) required the usage of 1024 nodes, mainly due to the increased memory requirements, as the use of velocity requires reading an additional 3D vector data set. An example of these combined data sets is shown in Figure \ref{f-front}, for the $1152^3$ output. 

Next, the sonic-scale filtered density is shown. Since this analysis highlights the gas parcels with velocity magnitude close to the speed of sound, it can be considered as a proxy for the shock locations, showing how ubiquitous and interconnected they are across the full box. This task was the least expensive from a computational viewpoint, as the shown data set remains the same throughout this part of the visualization, and only the camera position rotates. It was rendered in two solutions using 128 to 512 nodes, for balancing queue- and run- time on the system. An example from the $1152^3$ HD run, preferred for visual clarity, is reported in Figure \ref{f-sonic}, whereas in the movie the output of the full $10048^3$ HD run is shown.

\subsection{Time evolution: the $1008^3$ MHD case}
In a second series of renderings, we show the full time evolution of a MHD run, transitioning between the gas overdensity and the magnetic energy. For an evolution to have enough time coverage, a large number of data dumps have to be combined, and this was not feasible at full simulation resolution. In the MHD run with $1008^3$ grid resolution we saved $557$ data dumps, for a total disk storage of more than $15$ TB.

\begin{figure*}[t!]
      \centering
      \includegraphics[width=0.495\textwidth]{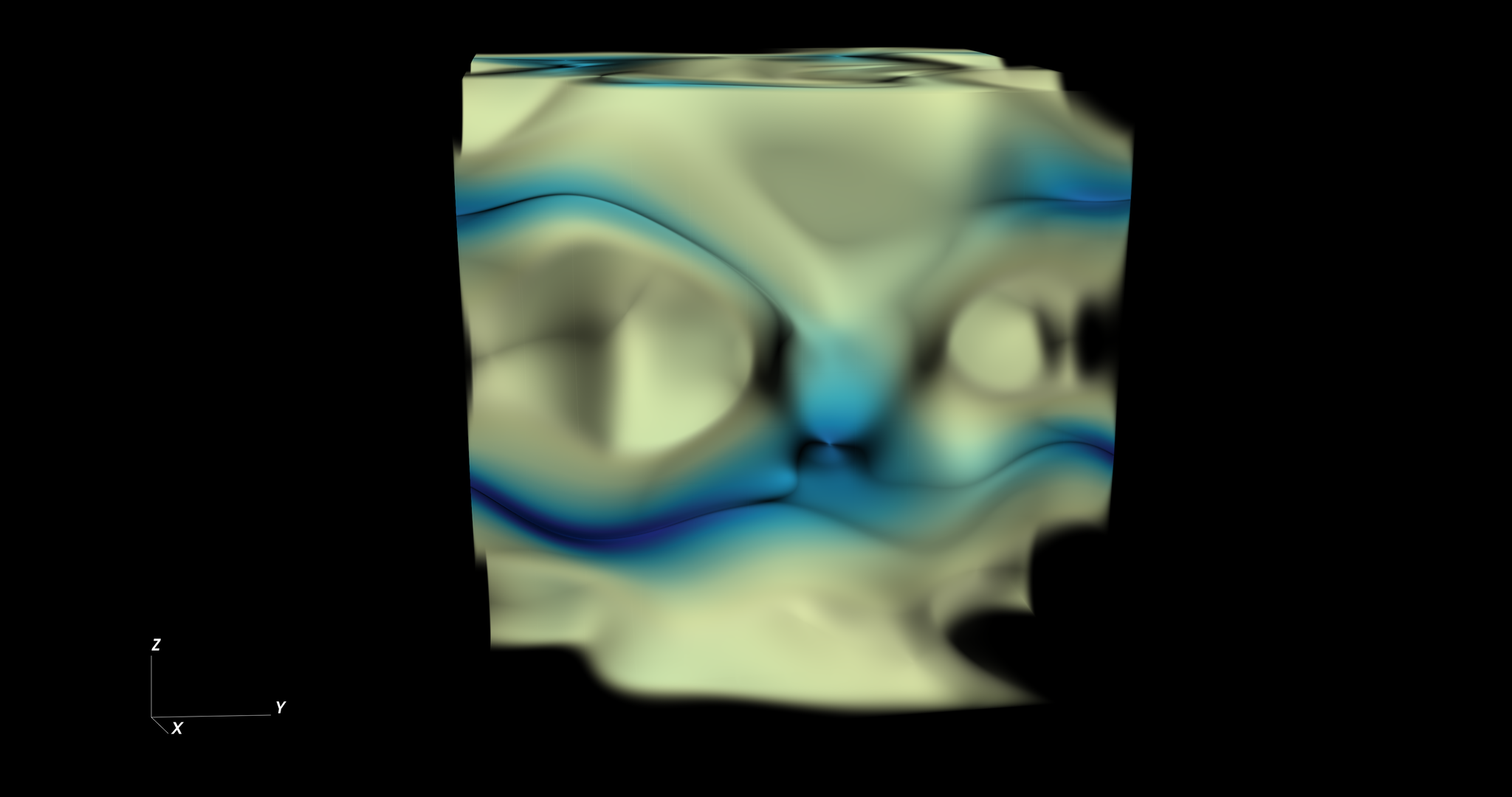}     
      \includegraphics[width=0.495\textwidth]{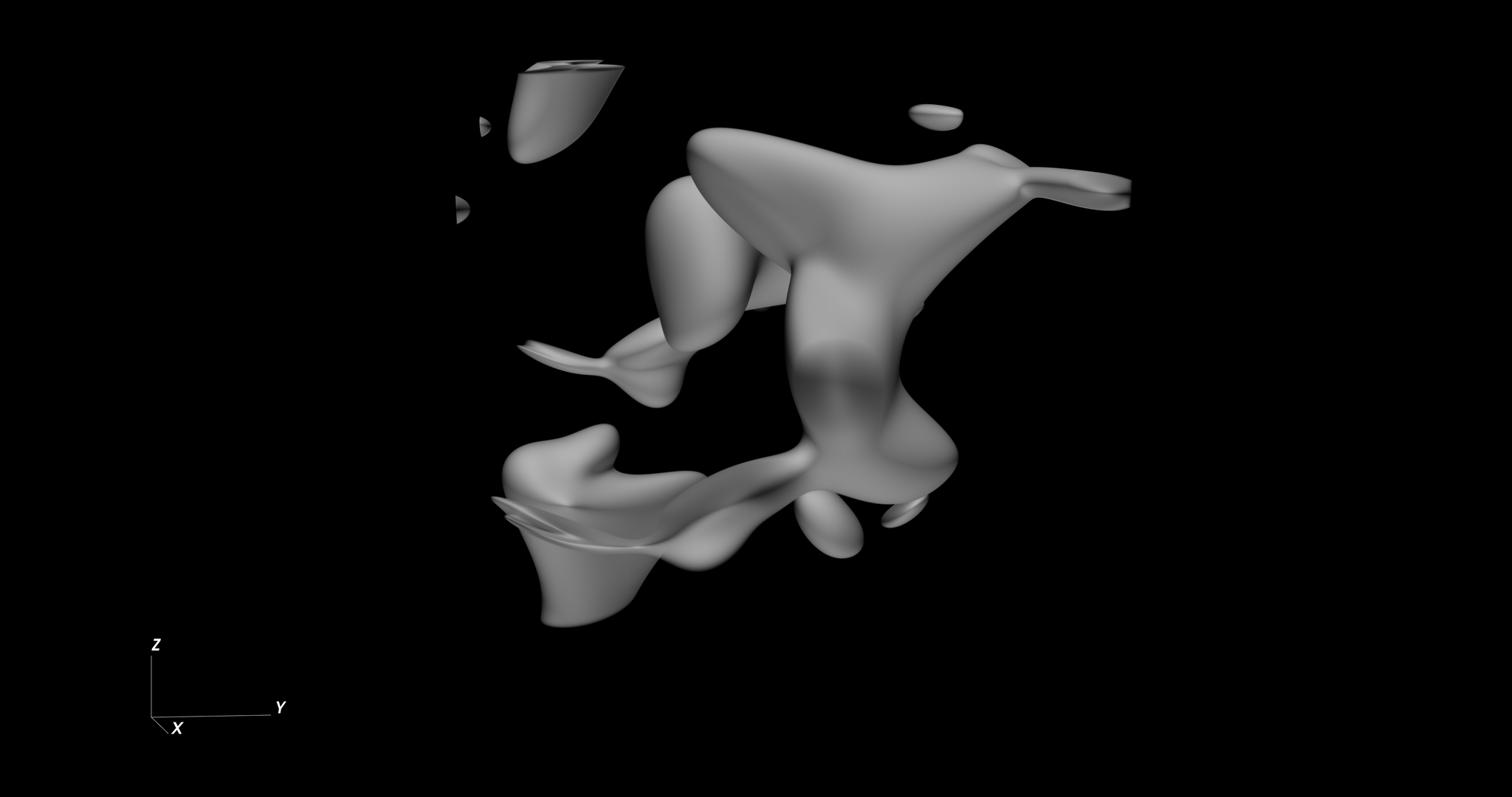}
      \includegraphics[width=0.495\textwidth]{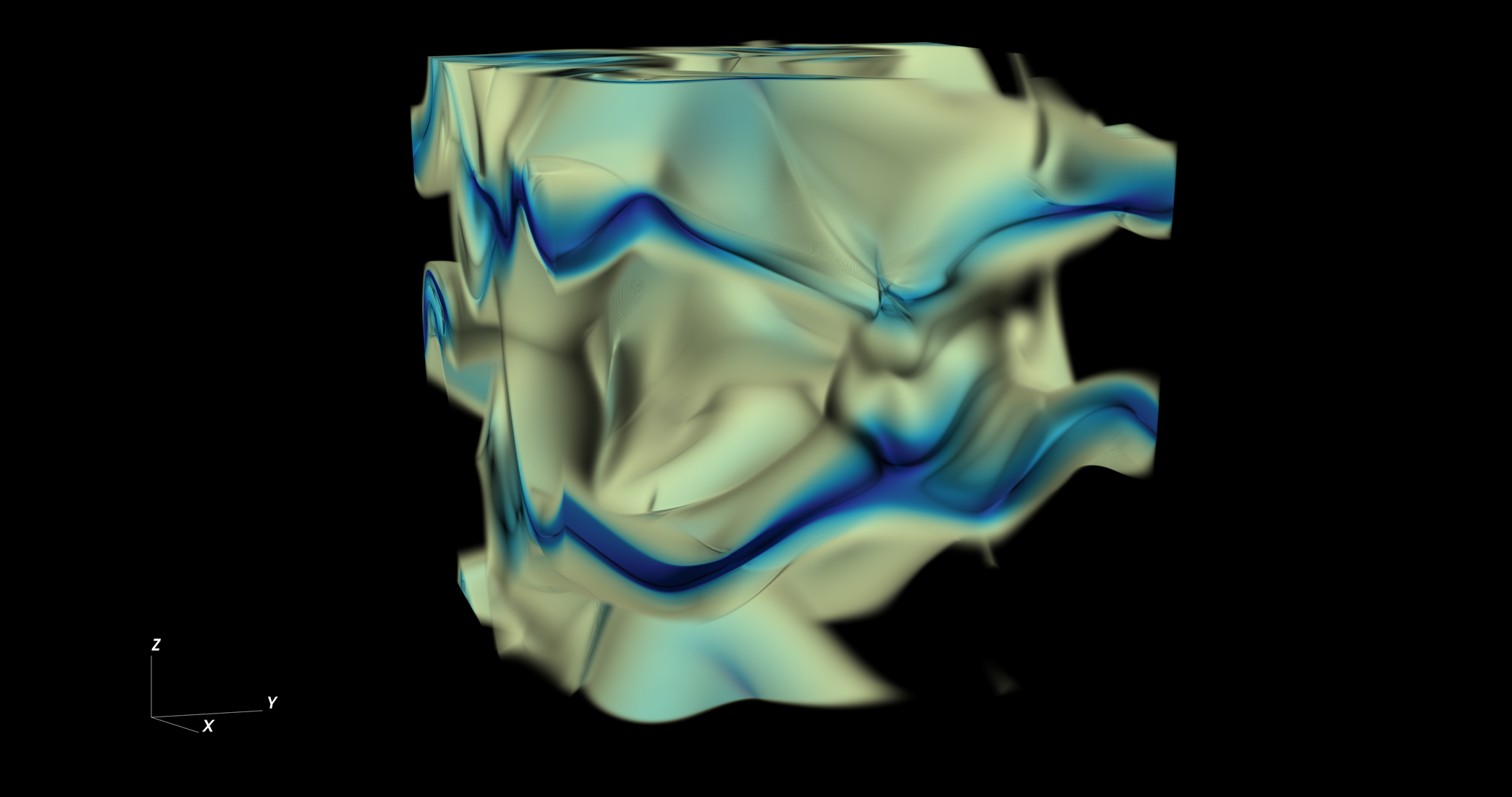}
      \includegraphics[width=0.495\textwidth]{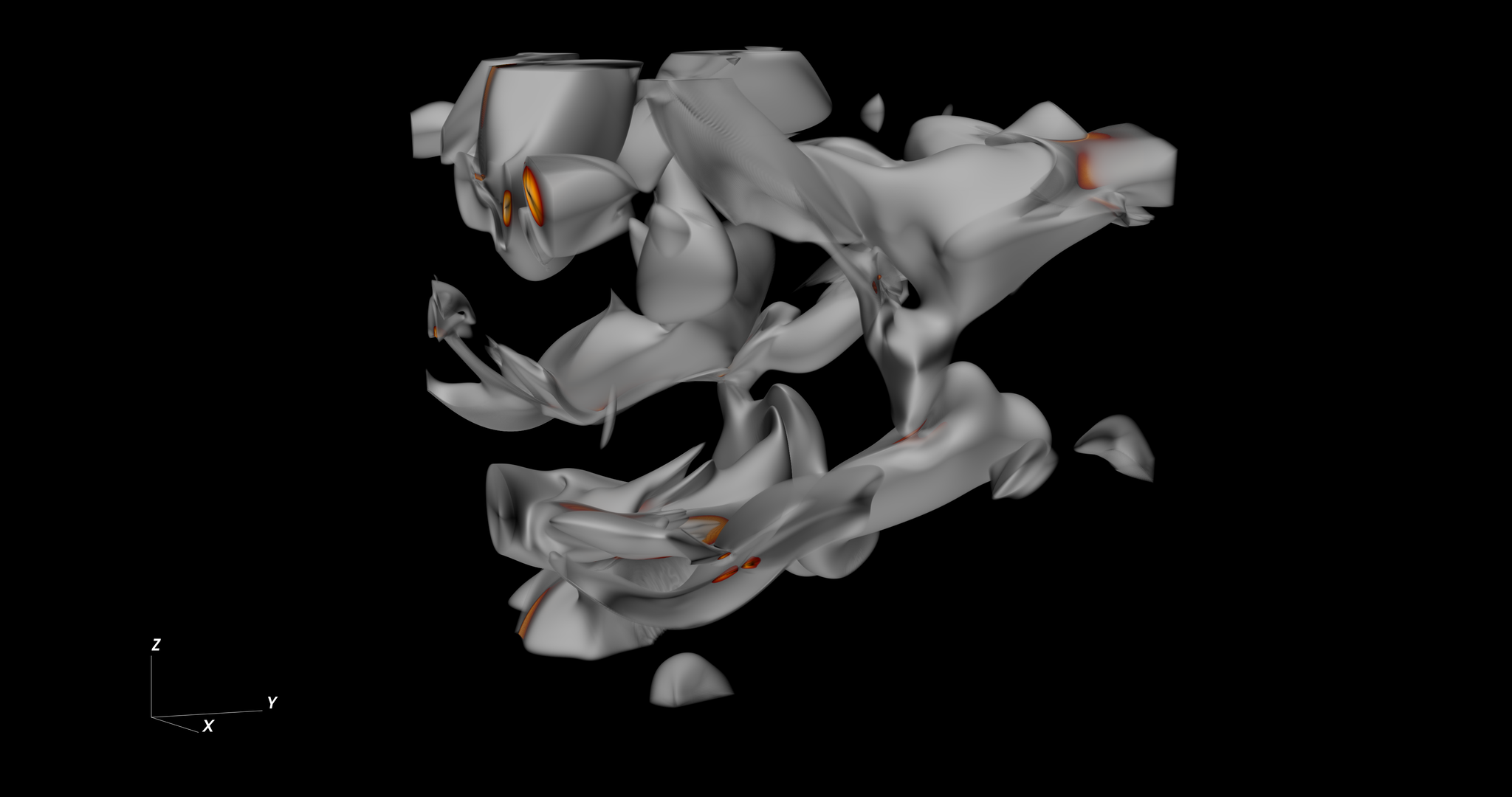}
      \includegraphics[width=0.495\textwidth]{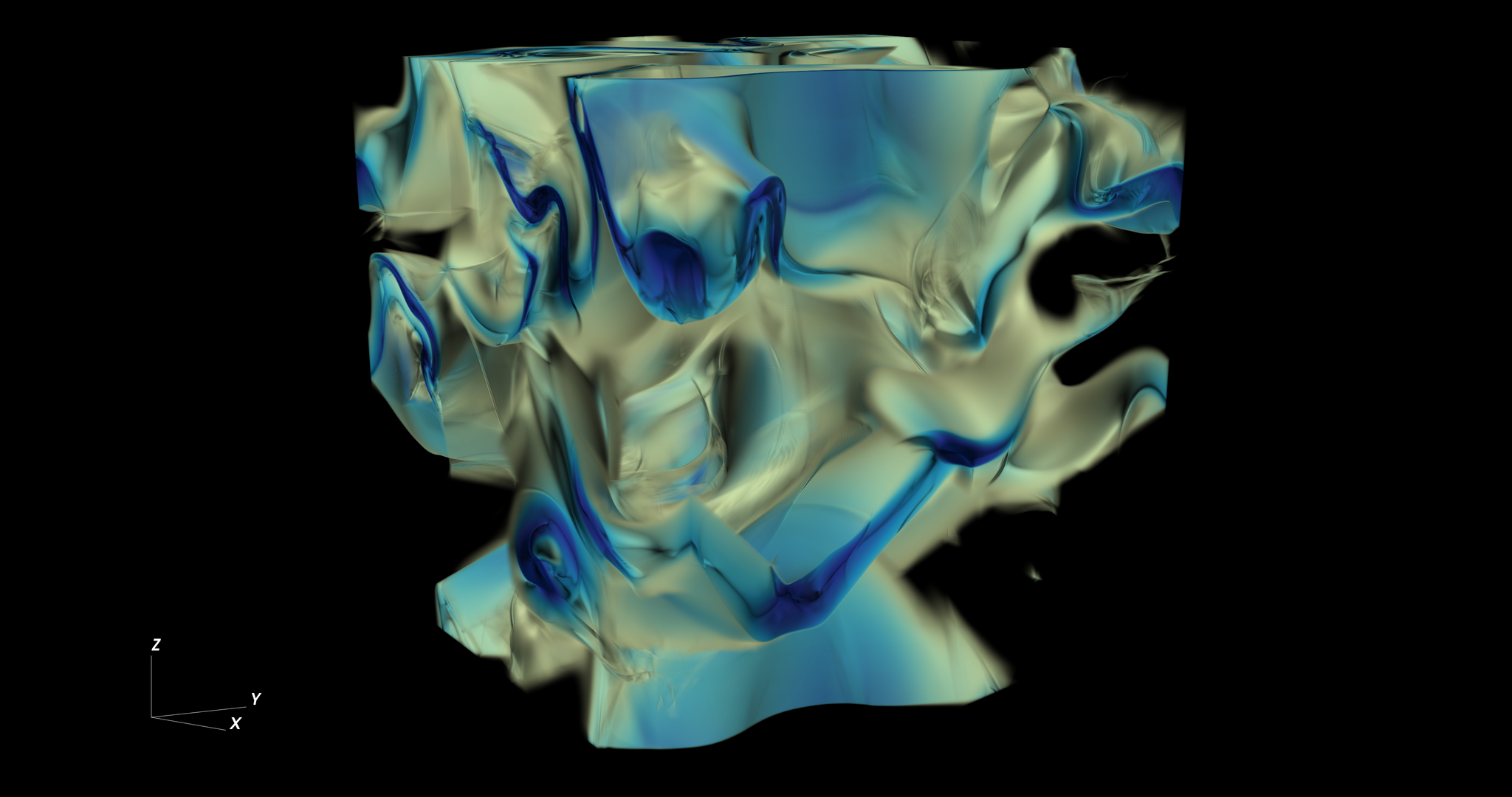}
      \includegraphics[width=0.495\textwidth]{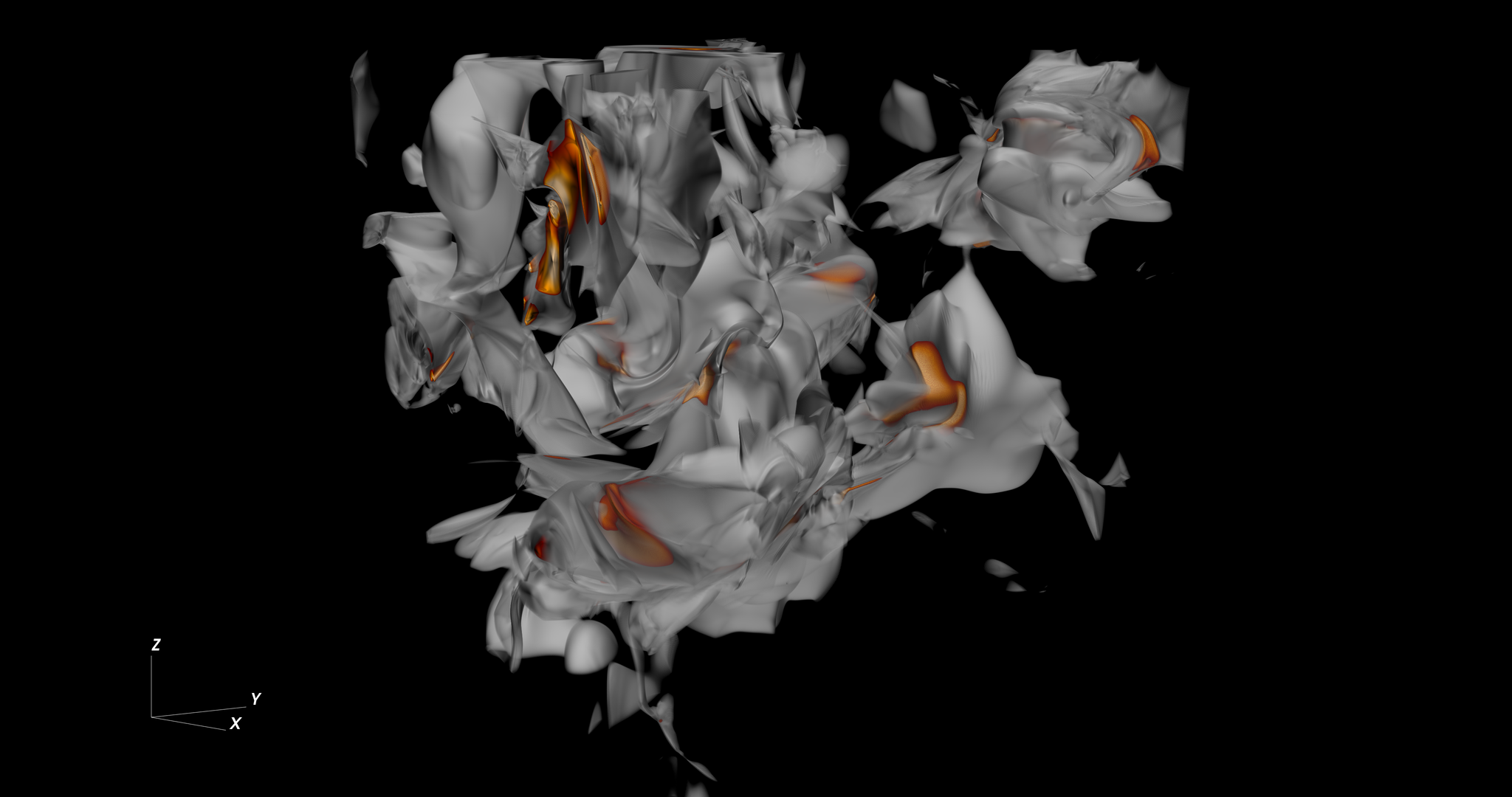}
      \includegraphics[width=0.495\textwidth]{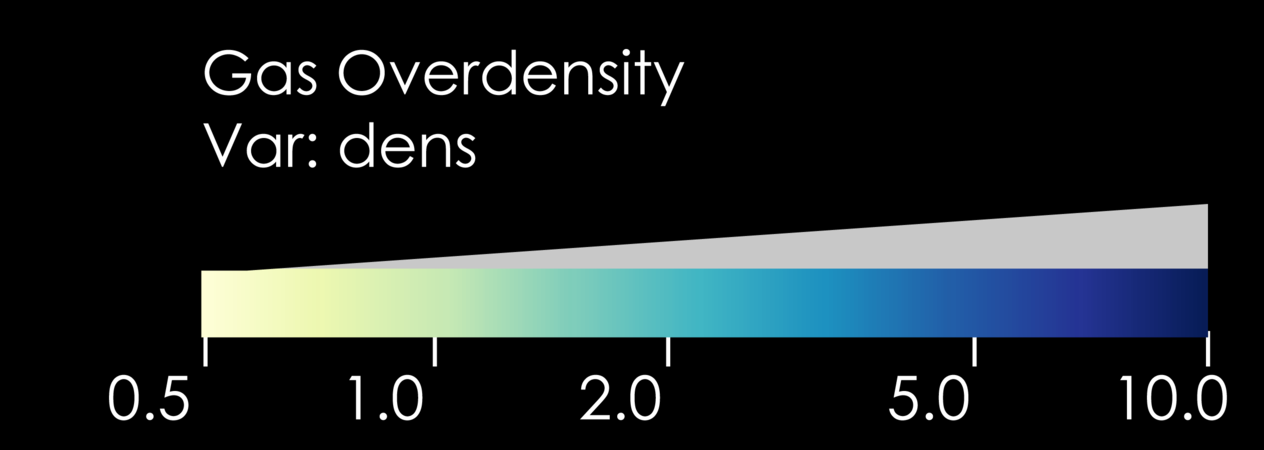}
      \includegraphics[width=0.495\textwidth]{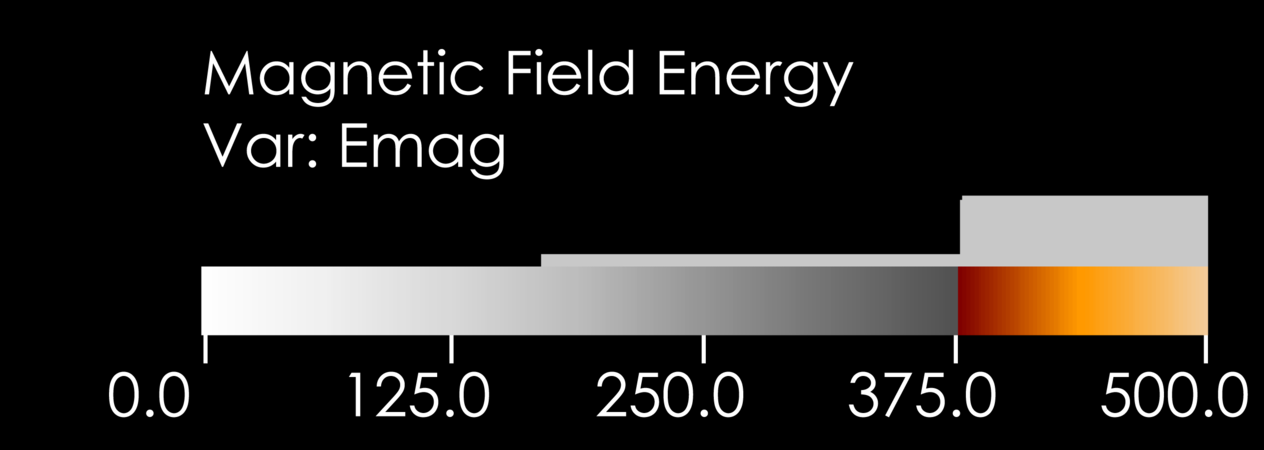}
      \caption{Density (left) and magnetic field energy (right) evolution in a $1008^3$ MHD run. The gas folding and the instabilities giving rise to turbulence are visible. By comparing the two data fields, the level of magnetic energy of the densest structures can be inferred. }
     \label{f-time}
 \end{figure*}

The strength of this visualization is to showcase the stretching and folding mechanisms, leading to the subsequent development of hydrodynamical instabilities 
and to the turbulent dynamo. A visual inspection shows that the magnetic field has a larger magnitude around the densest regions, forming filamentary structures in analogy to the ones in the density field (Figure \ref{f-time}). As for the rendering computation, while for the density a single node was sufficient, the magnetic field required four nodes.


\section{Concluding remarks}

With the increase of computing resources on the road to Exascale, it is getting more and more urgent to handle large data sets; the complexity of data analysis has become not smaller than the one of the simulations themselves. In our work we demonstrate that data visualization in the regime beyond $10$ TB is a HPC task that can make efficient use of high-end computing systems. These visualizations, besides having a remarkable visual impact, are extremely important for exploring large data sets and for steering further quantitative analysis.

\section*{Acknowledgements}
C.~F.~acknowledges funding provided by the Australian Research Council (Discovery Project DP170100603 and Future Fellowship FT180100495), and the Australia-Germany Joint Research Cooperation Scheme (UA-DAAD). We acknowledge high-performance computing resources provided by the Leibniz Supercomputing Centre and the Gauss Centre for Supercomputing (grants~pr32lo, pr48pi and GCS Large-scale project~10391), the Australian National Computational Infrastructure (grant~ek9) in the framework of the National Computational Merit Allocation Scheme and the ANU Allocation Scheme. The simulation software FLASH was in part developed by the DOE-supported Flash Center for Computational Science at the University of Chicago. VisIt is supported by the Department of Energy with funding from the Advanced Simulation and Computing Program and the Scientific Discovery through Advanced Computing Program.

\section*{References}
\bibliographystyle{elsarticle-num} 
\bibliography{mybib}

\end{document}